\documentstyle[aps,prl,epsfig,twocolumn]{revtex}
\draft
\newcommand{\pathfigs}{.}

\newcommand{\be}{\begin{equation}}
\newcommand{\ee}{\end{equation}}
\newcommand{\bea}{\begin{eqnarray}}
\newcommand{\eea}{\end{eqnarray}}
\newcommand{\bdm}{\begin{displaymath}}
\newcommand{\edm}{\end{displaymath}}

\newcommand{\bi}{\begin{itemize}}
\newcommand{\ei}{\end{itemize}}

\newcommand{\refkl}[1]{(\ref{#1})}

\newcommand{\sub}[1]{_{\rm #1}}

\newcommand{\tilQout}{\tilde{Q}_{\rm out}}

\newcommand{\rhomax}{\rho_{\rm max}}

\begin{document}

\tighten
\onecolumn
\twocolumn[\hsize\textwidth\columnwidth\hsize\csname @twocolumnfalse\endcsname
{\protect

\title{Explanation of observed features of self-organization
in traffic flow}
\author{Martin Treiber and Dirk Helbing}
\address{II. Institute of Theoretical Physics, University of Stuttgart,
         Pfaffenwaldring 57/III, D-70550 Stuttgart, Germany}

\date{\today}
\maketitle

\begin{abstract}
Based on simulations with the ``intelligent driver model'', a
microscopic traffic model, we explain the recently discovered 
transition from free over ``synchronized'' traffic to stop-and-go 
patterns [B.~S. Kerner, Phys. Rev. Lett. {\bf 81},  3797  (1998)].
We obtain a nearly quantitative agreement with empirical findings
such as the ``pinch effect'', the flow-density
diagram, the emergence of stop-and-go waves from
nonhomogeneous congested traffic, and the dimensions of their wavelength.
\end{abstract}
\vspace*{10mm}
} ]

During the last years, theoretical and empirical investigations have 
identified different possible mechanisms for a phase
transition from free traffic
to stop-and-go traffic on freeways. This includes deterministic 
\cite{KK-94,book,Wagner-96,Bando}
and stochastic mechanisms \cite{Nagel-S,Schreck95,Krauss-Wagner,Wolf} 
as well as effects of
inhomogeneities \cite{Lee,sync-Letter,Phase-prelim,TSG-science}. 
In contrast, Kerner has recently
described the detailled features of another transition to stop-and-go
patterns  \cite{Kerner-wide} developing from ``synchronized'' 
congested traffic \cite{Kerner-rehb96-2,Kerner-sync}
on German highways, 
which are compatible with
empirical findings on Dutch highways \cite{Helb-emp97}. 
\par
In the following,
we propose a quantitative explanation of these observations based on 
microsimulations with the ``intelligent driver model'' (IDM). 
In particular, we will show the possible coexistence of different
traffic states along the road behind an inhomogeneity of traffic flow.
It is, in upstream direction,
associated with the sequence ``homogeneous congested traffic'' 
\cite{sync-Letter,Phase-prelim}
(which, in a multilane model, is related to the observed
synchronization among lanes \cite{Lee,Vladi-98-prelim}) $\to$
``inhomogeneous congested traffic'' \cite{Phase-prelim}
(corresponding to the so-called ``pinch region'' \cite{Kerner-wide}) 
$\to$ ``stop-and-go
traffic'', while we have free traffic flow downstream of the
inhomogeneity.
\par
It will turn out that, in contrast to 
previously reported traffic phenomena, this phenomenon
relies on the existence of a sufficiently large density
region of convectively stable traffic, in which traffic flow is unstable,
but any perturbations are convected away in upstream direction.
Furthermore, one needs
a traffic model in which the resulting traffic flow
inside of fully developed traffic jams is much lower (nearly zero) than
in ``synchronized'' traffic.
In particular, without suitable modifications (see below), 
this is not satisfied by the 
traffic model discussed in Ref.~\cite{sync-Letter}.
We also point out that, although the IDM 
has a unique flow-density relation in equilibrium, it
reproduces the observed two-dimensional scattering
of flow-density data at medium vehicle densities 
\cite{Kerner-rehb96-2,Kerner-wide},
even without assuming a mixture of different vehicle types \cite{GKT-scatter}.
\par
The IDM is a continuous, deterministic model, in which the
acceleration of a vehicle $\alpha$ of length $l_\alpha$ at
position $x_{\alpha}(t)$ depends on its own velocity $v_{\alpha}(t)$ 
as well as the gap
$s_{\alpha}(t) = [x_{\alpha-1}(t)-x_{\alpha}(t) - l_{\alpha}]$ 
and the velocity difference 
$\Delta v_{\alpha}(t) = [v_{\alpha}(t) - v_{\alpha-1}(t)]$ to the 
vehicle $(\alpha-1)$ in front:
\begin{equation}
\label{IDMv}
\dot{v}_{\alpha} = a 
         \left[ 1 -\left( \frac{v_{\alpha}}{v_0} \right)^{\delta} 
                  - \left( \frac{s^*}{s_{\alpha}} \right)^2
         \right] \, .
\end{equation}
According to this formula,
the acceleration on a free road (meaning $s_{\alpha}\to\infty$) is given by
$a[1-(v_{\alpha}/v_0)^{\delta} ]$, where $a$ is the maximum acceleration
and $v_0$ the desired velocity. The exponent $\delta$ is typically
between 1 and 5. It allows to 
describe that the realistic acceleration behavior of drivers lies
between a constant acceleration $a$ up to their desired velocity $v_0$
($\delta \to \infty$) and an exponential acceleration behavior
($\delta = 1$).

The braking term $-a(s^*/s_{\alpha})^2$
depends Coulomb-like on the gap $s_{\alpha}$, as it is the case
for the braking term of the microscopic Wiedemann model \cite{Wiedemann}.
Therefore, the acceleration term is negligible, if the gap $s_\alpha$
drops considerably below the
``effective desired distance'' $s^*$. With the relation
\be
\label{dxstar}
s^*(v_{\alpha}, \Delta v_{\alpha}) = 
  s_0 + \mbox{max} \left( v_{\alpha} T 
     + \frac{v_{\alpha} \Delta v_{\alpha} }  {2\sqrt{a b}}, 0 \right) \, ,
\ee
it is constructed in a way that
drivers keep a minimum ``jam distance'' $s_0$ to a standing vehicle,
plus an additional safety distance $v_\alpha T$,
where $T$ is the safe time headway
in congested but moving traffic.

The nonequilibrium term proportional to $\Delta v_{\alpha}$
\cite{book,Wolf}
reflects  an ``intelligent'' braking strategy, according to which
drivers restrict their deceleration to $b$ in ``normal''
situations (e.g., when approaching 
standing or slower vehicles from
sufficiently large distances),
but they brake harder when the situation becomes more critical,
i.e., when the anticipated ``kinematic deceleration'' $(\Delta
v)^2/(2s_\alpha)$, which is necessary to avoid a collision with
a uniformly moving leading vehicle ($\dot{v}_{\alpha-1}=0$),
exceeds $b$.
Notice that the acceleration $a$ is typically lower 
than the desired deceleration $b$,
and that both acceleration parameters do not
influence the equilibrium flow-density relation (``fundamental diagram'').
Since it turns out that neither multilane effects nor different
types of vehicles are relevant in the context of this study, we
have assumed identical ``driver-vehicle units'' 
characterized by the realistic parameters
$v_0=120$ km/h, 
$\delta=4$, 
$a=0.6$ m/s$^2$, 
$b=0.9$ m/s$^2$, 
$s_0=2$ m, and
$T=1.5$ s, apart from a localized change of $v_0$ or $T$ (see below). 
For the vehicle length we use
$l=5$ m, but this value does not affect the dynamics.

\begin{figure}

\vspace*{-30\unitlength}

\begin{center}
   \includegraphics[width=150\unitlength]{\pathfigs/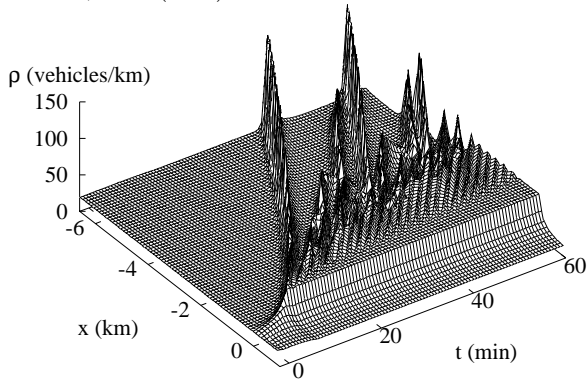}
\end{center}

\caption[]{
\label{fig_rho3D}
Spatiotemporal density plot illustrating the breakdown to ``synchronized''
traffic (smooth region of high density)
near an inhomogeneity, and showing stop-and-go waves
emanating from this region.
Traffic flows in positive $x$-direction.
The inhomogeneity corresponds to an increased safe time headway $T$
between $x=0$ km and $x=0.3$ km, reflecting more careful driving 
(see main text). Downstream of the inhomogeneity, vehicles accelerate
into free traffic.
}

\end{figure}
\begin{figure}

\vspace*{0\unitlength}

\begin{center}
   \hspace*{-12\unitlength}
   \includegraphics[width=80\unitlength]{\pathfigs/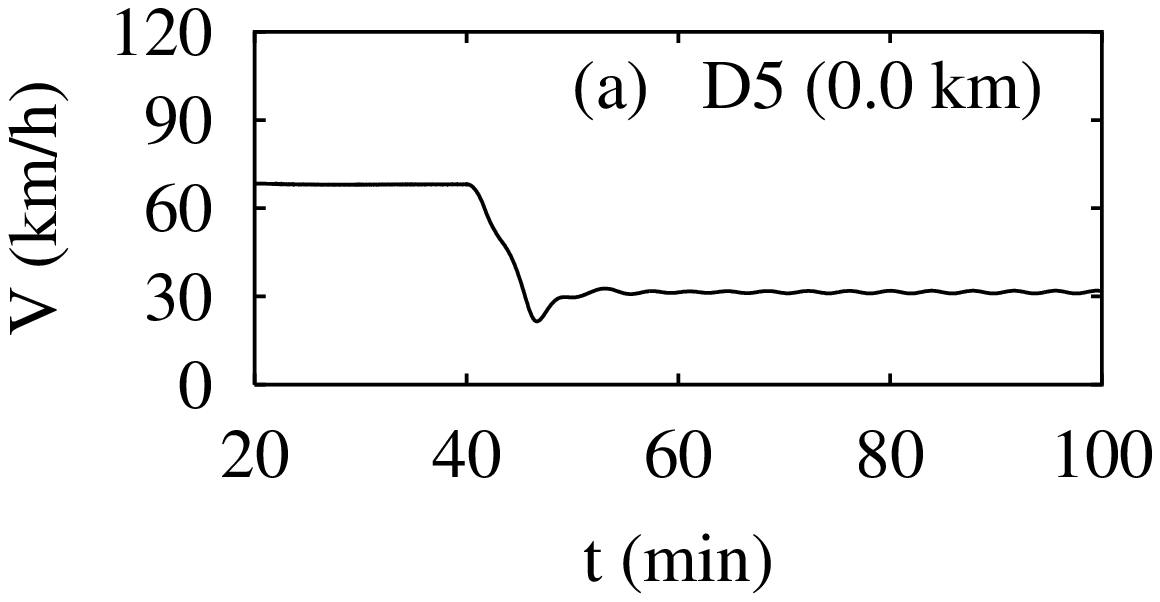}  
   \hspace*{-12\unitlength}
   \includegraphics[width=80\unitlength]{\pathfigs/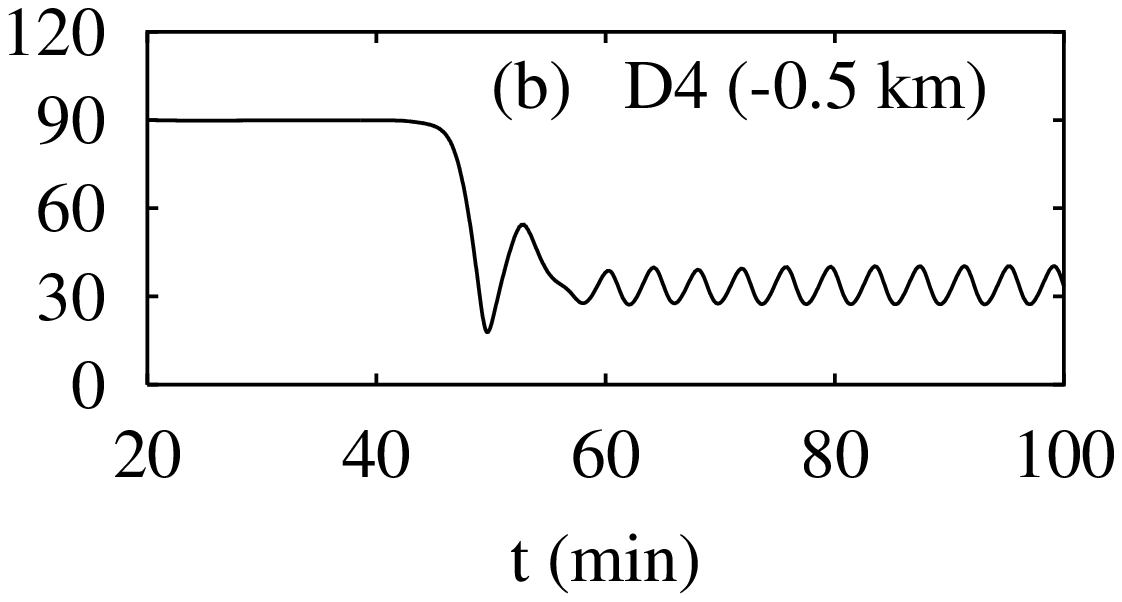}  
   \\[0\unitlength]
   \hspace*{-12\unitlength}
   \includegraphics[width=80\unitlength]{\pathfigs/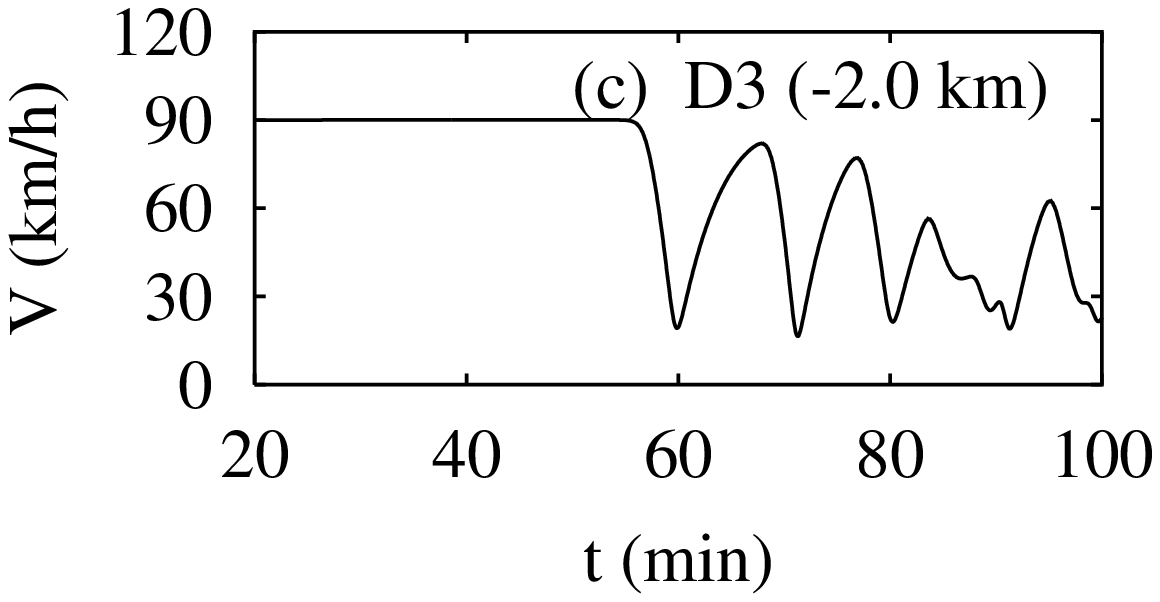}  
   \hspace*{-12\unitlength}
   \includegraphics[width=80\unitlength]{\pathfigs/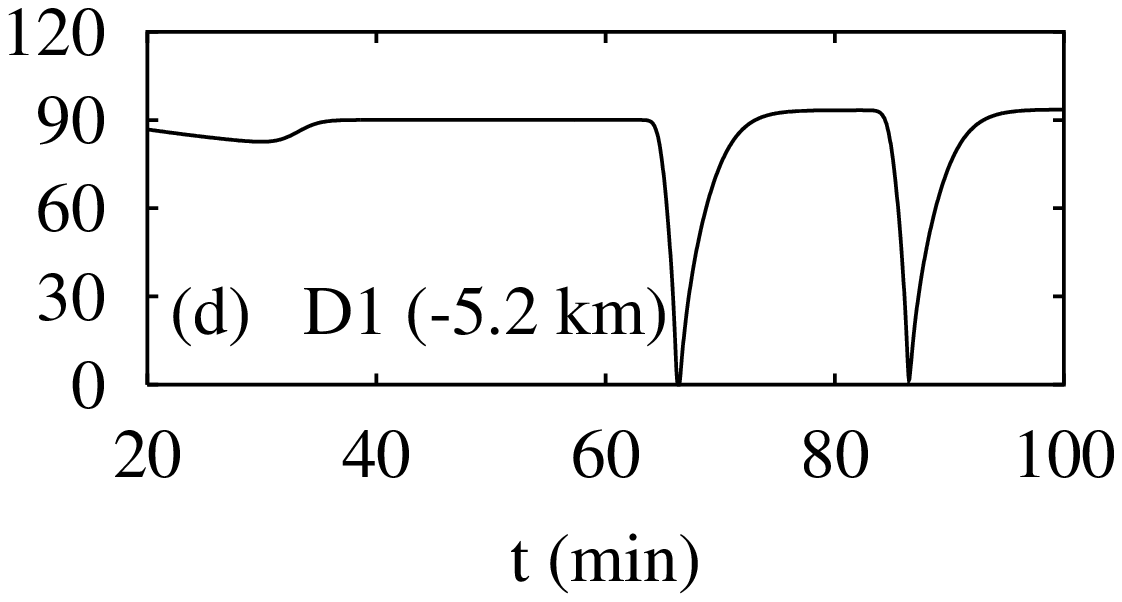} 
\end{center}
\caption[]{
\label{fig_vt}
Temporal evolution of the average velocity determined from 
the individual velocities of the vehicles that pass cross sections
of the freeway during one-minute intervals at
the four ``detector'' positions D1, D3, D4
(upstream of the inhomogeneity), and D5 (at the inhomogeneity).
The naming of the detectors and their
distances with respect to the inhomogeneity are the same as
in Ref.~\protect\cite{Kerner-wide}.
The inhomogeneity  is realized by a drop of the desired
velocity
(see the main text).
(a) Breakdown to homogeneous synchronized traffic around
$t=45$ min. 
(b) Oscillating synchronized traffic (``pinch region''),
(c) developing stop-and-go waves, and
(d) resulting traffic jams. 
}
\end{figure}
\begin{figure}

\begin{center}
   \includegraphics[width=130\unitlength]{\pathfigs/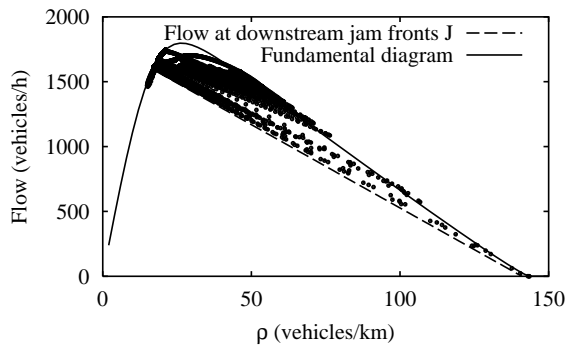} 
\end{center}

\caption[]{
\label{fig_fund}
Flow-density data (symbols)
calculated from the microscopic simulation  
at the four detector locations displayed in Fig.~\protect\ref{fig_vt},
and in addition at $x=-3.7$ km (detector D2) and at 
$x=1.2$ km (D6).
Also shown is the equilibrium flow-density relation (solid line),
and the flow-density relation at the downstream fronts of
fully developed jams (dashed).
}
\end{figure}
\begin{figure}

\vspace*{-30\unitlength}
\begin{center}
   \includegraphics[width=150\unitlength]{\pathfigs/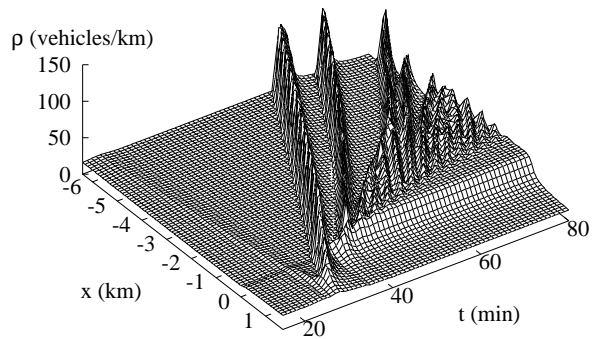} 
\end{center}

\caption{
\label{fig_macrho3D}
Spatiotemporal evolution of the traffic density according to the
gas-kinetic-based traffic model \protect\cite{GKT}.
The assumed inhomogeneity of traffic flow comes from an on-ramp 
of length 200 m with an inflow of 220 vehicles per hour
and freeway lane. The inflow to the main road is 
1570 vehicles per hour and lane, and
the breakdown of traffic flow is triggered
by a perturbation $\Delta Q(t)$ of the inflow with a flow peak of 125
vehicles per hour and lane (see main text). 
The assumed model parameters are $V_0=$ 120 km/h, 
$T=1.5$ s,
$\tau=30$ s,
$\rhomax=$ 120 vehicles/km, and
$\gamma=1.2$, while
the parameters for the variance prefactor \protect\cite{GKT} 
$A(\rho) = A_0 + \Delta A \{\tanh [ (\rho-\rho_{\rm c})/\Delta \rho] + 1\}$
are
$A_0=0.008$,
$\Delta A = 0.02$,
$\rho_{\rm c} = 0.27 \rhomax$, and
$\Delta \rho = 0.1 \rhomax$.
In order to have a large region of linearly unstable but convectively
stable traffic, we introduced
a ``resignation effect'', i.e.\ a density-dependent reduction of the
desired velocity $V_0$ to
$V'_0(\rho) = V_0 - \Delta V / \{1 + \exp[(\rho'_{\rm c}-\rho)/\Delta \rho']\}$
with
$\Delta V = 0.9 V_0$,
$\rho'_{\rm c} = 0.45 \rhomax$, and
$\Delta \rho = 0.1 \rhomax$.
}
\end{figure}

We simulated an open freeway section of 20 kilometer length 
for time intervals up to 120 minutes, of which we display 
the most interesting parts only.
In addition, we assumed an inhomogeneity of traffic flow that will
be responsible for the transition from free to congested traffic,
as described in Refs.~\cite{sync-Letter,Phase-prelim}. 
However, as pointed out by Kerner \cite{Kerner-wide}, the self-organized
patterns observed by him are not restricted to the vicinity
of on-ramps. We will confirm this by
simulating different kinds of inhomogeneities 
(see Figs.~\ref{fig_rho3D} through \ref{fig_fund}) and comparing them
with the injection of vehicles at on-ramps (see Fig.~\ref{fig_macrho3D}).


In Figure~\ref{fig_rho3D},
we have assumed an inhomogeneity 
corresponding to a freeway section where people drive
more carefully. This was modelled by setting the desired 
time headway from $T=1.5$ s to  $T=1.75$ s 
between $x=0$ km and $x=0.3$ km. In the simulations of
Figs.~\ref{fig_vt} and \ref{fig_fund}, we have 
reduced the desired velocity from
$v_0 = 120$ km/h to $v_0 = 80$ km/h in the same region.
As initial conditions, we assumed 
homogeneous free traffic in equilibrium
at a flow of $Q_0 = 1670$ vehicles/h
(Figs.~\ref{fig_rho3D} through \ref{fig_fund})
or 1570 vehicles/h (Fig. \ref{fig_macrho3D}). 
The actual initial conditions, however, are
only relevant for a short time interval.
At the upstream boundary, we assume that vehicles enter the freeway 
uniformly at a rate $Q\sub{in}(t) = Q_0 + \Delta Q(t)$ and drive with a
velocity corresponding to free traffic in equilibrium.
While in the simulation of Fig.~\ref{fig_rho3D}, the breakdown of traffic
flow is caused by exceeding the static freeway
capacity at the inhomogeneity \cite{Phase-prelim}, in Figs.~\ref{fig_vt} and
\ref{fig_fund} it is triggered by a triangularly shaped
perturbation $\Delta Q(t)$ of the inflow \cite{sync-Letter,Phase-prelim}
between $t=10$ min and $t=20$ min with a maximum of 
200 vehicles per hour and lane at $t=15$ min.
Furthermore, we used the ``absorbing'' downstream boundary condition
$\dot{v}_{\alpha} = 0$. 
To minimize simulation time, we integrated Eq. \refkl{IDMv} 
with a simple Euler scheme using a coarse
time discretization of $\Delta t =  0.4$ s,
and translated the vehicles in each step according to 
$x_{\alpha}(t+\Delta t) = x_{\alpha}(t) + v_{\alpha} \Delta t
+ \frac{1}{2} \dot{v}_{\alpha}(\Delta t)^2$. However, smaller values
of $\Delta t$ yielded nearly indistinguishable  results.

Figure \ref{fig_rho3D} gives a representative 
overview of the simulation result
by means of a spatiotemporal density plot.
The resulting sequence of transitions
is essentially the same as observed \cite{Kerner-wide}:
After 10 minutes of free traffic, traffic breaks
down near the inhomogeneity, resulting in homogeneous
congested traffic at this location, that persists over a long time. 
Upstream of the inhomogeneity, small oscillations develop that travel
further upstream and grow to stop-and-go waves of relatively short
wavelengths
(about 0.8 km). Finally, these waves either dissolve or merge to a few 
``wide jams'' (in which traffic comes to a standstill)
with typical distances of 2 km up to 5 km between them.
Once the jams have formed, they persist and propagate upstream
at a constant propagation velocity without further changes of their
shape. No new clusters develop between the jams.

To compare our simulation results directly with the empirical data
published by Kerner \cite{Kerner-wide}, we investigated the temporal evolution
of the average velocity at six subsequent locations D1 through D6 that
had the same distances with respect to the inhomogeneity as
in Ref.~\cite{Kerner-wide}
(see Fig.~\ref{fig_vt}). 
The detector positions D1 through D4 are upstream 
of the inhomogeneity, D5 is directly at the inhomogeneity, 
and D6 is downstream of it.
In contrast to the simulation of Fig. \ref{fig_rho3D}, the
capacity drop at the inhomogeneity is so weak that free
traffic is metastable in the overall system. 
At the inhomogeneity (D5), one observes homogeneous congested
(``synchronized'') traffic, at D4 one sees small oscillations that,
around D3, develop to stop-and-go waves of larger amplitude, 
and finally to jams (at D1 and D2). In the downstream
direction, the congested traffic dissolves to free traffic (D6).
Apart from irregularities in the measured data due to fluctuations, 
the curves are in (semi-)quantitative
agreement with Kerner's empirical findings \cite{Kerner-wide}.

We also plot one-minute data of the six ``detectors'' in a
flow-density diagram (Fig.~\ref{fig_fund}), together with the
equilibrium flow-density relation $Q_{\rm e}(\rho)$ (``fundamental
diagram''). The lower boundary of the data points at medium densities
corresponds to the flow-density relation belonging to the downstream front
of a fully developed jam.
In agreement with observations, this line is the same for all jams and 
corresponds to a unique propagation velocity of their downstream fronts.
The twodimensional region of points above this line relate
to congested traffic at D3, D4, and D5.


To understand this scenario,
we need some basic results about
the stability of homogeneous traffic with respect to localized 
perturbations. Typically, there are
four ``critical'' densities $\rho_{{\rm c}i}$ with
$\rho\sub{c1} < \rho\sub{c2} < \rho\sub{c3} < \rho\sub{c4}$ and
the following properties \cite{Kerner-dipole,GKT}:
For low  and very high densities
($\rho<\rho\sub{c1}$ or 
$\rho>\rho\sub{c4}$), traffic is stable with
respect to arbitrary perturbations, while 
for $\rho\sub{c2} < \rho < \rho\sub{c3}$, it is linearly unstable.
In the two density ranges in between,
homogeneous traffic is unstable
only with respect to perturbations exceeding a certain critical
amplitude $\Delta \rho\sub{cr}(\rho)$
(``metastability''). Furthermore, there exists a range 
$\rho\sub{cv} < \rho < \rho\sub{c3}$ with
$\rho\sub{cv} > \rho\sub{c2}$, where traffic is linearly
unstable, but convectively stable, i.e., all perturbations grow, but
they are eventually convected away in upstream direction
\cite{Cross-Hohenberg}. Actually, this range can be very large.
For the IDM
with the parameters used here, we have $\rho\sub{cv}=50$ vehicles/km and
$\rho\sub{c3}=100$ vehicles/km.

Let us assume that the inflow $Q\sub{in}$ has a value larger than
$Q\sub{c1}$, and that a phase transition from
free to congested traffic has occurred at the inhomogeneity.
This breakdown  to congested traffic, which we identify with
``synchronized traffic'', here 
\cite{sync-Letter,Phase-prelim,Vladi-98-prelim,GKT-scatter},
was explained in \cite{sync-Letter}. 
Since it 
is connected with a
drop of the {\em effective} freeway capacity 
to the self-organized outflow $\tilQout$ 
from ``synchronized''
traffic (ST), the region of congested traffic will grow in
upstream direction until the inflow 
$Q\sub{in}(t) = Q_{\rm e}(\rho\sub{in}(t))$ 
to the freeway falls below the synchronized
flow $Q\sub{ST}$ in the congested region behind
the inhomogeneity. 
Without an on-ramp flow $Q\sub{ramp}$ per freeway lane, we have
$Q\sub{ST} = \tilQout$, otherwise it is
$Q\sub{ST} = (\tilQout - Q\sub{ramp})$ \cite{sync-Letter}. 
In contrast to the empirical findings explained
in Ref.~\cite{sync-Letter}, the synchronized flow 
must be so high, here, 
that it is linearly unstable, which implies 
$Q\sub{ST}>Q_{\rm e}(\rho\sub{c3})$.
Therefore, small perturbations will grow to larger
oscillations, which propagate in upstream direction faster than the
congested region grows. When the oscillations reach the metastable region of
free traffic upstream of the inhomogeneity, oscillations with
an amplitude below the critical amplitude 
$\Delta \rho\sub{cr}(\rho\sub{in})$ will eventually disappear, the remaining
ones will continue to grow until they are fully developed traffic jams.
We point out that 
a sequence of such jams is sustained, because the propagation
velocity $v_{\rm g}$ and the outflow 
$Q\sub{out} \approx Q_{\rm e}(\rho\sub{c1})$
from jams are characteristic constants 
\cite{Kerner-rehb96,Kerner-asympt}. 
Finally, if the inhomogeneity is such that
the synchronized flow is convectively stable, i.e.,
$Q\sub{ST}<Q\sub{cv}= Q_{\rm e}(\rho\sub{cv})$, the perturbations
cannot propagate downstream (in contrast to small perturbations in
free traffic). Hence, the front of
dissolving traffic is smooth, then, and  
a small region of homogeneous congested traffic forms near the inhomogeneity
[Figs.~\ref{fig_rho3D}, \ref{fig_vt}(a), and \ref{fig_macrho3D}].

In 
Fig.~\ref{fig_fund}, 
the flow-density relation of the downstream front of a fully developed jam
(where the velocity is zero) corresponds 
to a straight line, the slope of which is 
$v_{\rm g}$ \cite{Kerner-rehb96}.
Notice that this line (which in Fig. \ref{fig_fund}
and in Ref.~\cite{Kerner-wide} is labelled by ``$J$'')
lies considerably below the equilibrium curve $Q_{\rm e}(\rho)$. On the
other hand, traffic flow in the region of
oscillating congested traffic is nearly in equilibrium as long as
the oscillations are small. As the oscillations grow,
the data points gradually approach the line $J$.
This explains the observed twodimensionality of the 
congested part of the flow-density diagram. We point out 
that 
mixtures of different types of driver-vehicle units 
lead to to further effects contributing to an even wider scattering of 
flow-density points in the congested regime \cite{GKT-scatter}.
%
%


In summary, we have shown that the emergence
of stop-and-go waves out of synchronized traffic, their coexistence,
and the twodimensional scattering of data in the
congested part of the flow-density diagram 
can be explained in the framework of ``standard'' traffic models
that have a unique  equilibrium  flow-density relation.
The necessary conditions
are a metastability of traffic flow, a flow inside of
traffic jams that is much lower than in synchronized congested traffic,
and a sufficiently
large density regime of linearly unstable traffic flow that is
convectively stable. 
If ``synchronized'' traffic is linearly unstable and free traffic
upstream is metastable, upstream-moving perturbations
will grow and, when their amplitudes became large enough,
eventually form stop-and-go waves. 
To maintain this mechanism, the region of synchronized traffic 
behind the inhomogeneity must persist (i.e. it must not dissolve to
stop-and-go waves), which is only the case if it is convectively stable.
The twodimensionality of the congested branch of the flow-density diagram
originates from the fact that the
nonhomogeneous congested states are not in equilibrium.

The phenomenon should be widespread  since it is
triggered at relatively small inhomogeneities
whenever traffic flow $Q_{\rm in}(t)$ exceeds a certain threshold $Q\sub{c1}$.
Note that it can be also simulated with
other traffic models like the macroscopic gas-kinetic-based model
\cite{GKT}, if ``frustration effects'' are additionally taken into account
(see Fig.~\ref{fig_macrho3D}).


The authors are grateful for financial support by the BMBF (research
project SANDY, grant No.~13N7092) and by the DFG 
(Heisenberg scholarship He~2789/1-1).


\bibliographystyle{/home/TeTeX/inputs/revtex/prsty}
\bibliography{/home/treiber/tex/bibtex/traffic}





\end{document}